\documentclass[runningheads]{llncs}
\usepackage{graphicx}

\usepackage[T1]{fontenc} 

\usepackage[frozencache,cachedir=.]{minted}
\usepackage[dvipsnames]{xcolor}
\definecolor{bg}{rgb}{0.95,0.95,0.95}
\usepackage{tikz}
\usetikzlibrary{chains,shadows.blur, quotes, positioning}
\usetikzlibrary{arrows, automata, shapes.multipart, chains, positioning, fit, graphs, spy, shapes, backgrounds, patterns,decorations.pathmorphing}
\usetikzlibrary{arrows.meta,bending,positioning}
\usepackage{tabularx}
\usepackage{amsmath}
\allowdisplaybreaks
\usepackage{bbold} 
\usepackage{MnSymbol} 
\usepackage{simplebnf}
\usepackage{stmaryrd}
\usepackage{algorithm}
\usepackage[noend]{algpseudocode}
\usepackage{tcolorbox}
\usepackage{subcaption}
\tcbuselibrary{skins,breakable, listings}
\tcbset{breakable, enhanced}
\newcommand\blockdistance{1cm}
\usepackage{tikz, pgfplots}

\definecolor{matlab_color_1}{RGB}{000, 114, 189}
\definecolor{matlab_color_2}{RGB}{217, 083, 025}
\definecolor{matlab_color_3}{RGB}{237, 177, 032}
\definecolor{matlab_color_4}{RGB}{126, 047, 142}
\definecolor{matlab_color_5}{RGB}{119, 172, 048}
\definecolor{matlab_color_6}{RGB}{077, 190, 238}
\definecolor{matlab_color_7}{RGB}{162, 020, 047}

\newtcbox{\innerbox}{nobeforeafter, colupper=white,colback=black!75!white,colframe=black, tcbox raise base, nobeforeafter,top=0pt,bottom=0pt,left=0mm,right=0mm,toprule=0.3mm,bottomrule=0.3mm, boxsep=0.5mm}

\usepackage{accents}
\newcommand{\ubar}[1]{\underaccent{\bar}{#1}}

\usepackage{wrapfig}
\usepackage[mode=buildnew]{standalone}
\usepackage{tikz}
\usepackage{pgf}
\usepackage{epsfig,psfrag}
\usepackage{enumitem}
\usetikzlibrary{arrows, automata, shapes.multipart, chains, positioning, fit, graphs, spy, shapes, backgrounds, patterns,decorations.pathmorphing}
\usetikzlibrary{arrows.meta,bending,positioning}
\newcommand\ppbb{path picture bounding box}
\usepackage{verbatim}

\usepackage[
    group-digits=integer, group-minimum-digits=4, 
    list-final-separator={, and }, add-integer-zero=false,
    free-standing-units, unit-optional-argument, 
    binary-units,
    detect-weight=true, detect-inline-weight=math, 
    round-mode=figures, round-precision=3, 
    ]{siunitx}[=v2]

\pgfplotsset{compat=1.18}
\begin{document}
\title{Interval Analysis in Industrial-Scale BMC Software Verifiers: A Case Study}
%
\author{Rafael Sá Menezes\inst{1,3}\orcidID{0000-0002-6102-4343} \and 
Edoardo Manino\inst{1}\orcidID{   0000-0003-0028-5440} \and 
Fedor Shmarov\inst{2}\orcidID{0000-0002-3848-451X} \and
Mohannad Aldughaim\inst{1,4}\orcidID{0000-0003-1708-1399} \and
Rosiane de Freitas\inst{3}\orcidID{0000-0002-7608-2052} \and
Lucas C. Cordeiro\inst{1,3}\orcidID{0000-0002-6235-4272}
}

\authorrunning{Rafael Sá Menezes et al.}
%
\institute{University of Manchester, UK \and
Newcastle University, UK \and
Universidade Federal do Amazonas, Brazil \and
King Saud University, Saudi Arabia}

\maketitle              
\begin{abstract}
Bounded Model Checking (BMC) is a widely used software verification technique. Despite its successes, the technique has several limiting factors, from state-space explosion to lack of completeness. Over the years, interval analysis has repeatedly been proposed as a partial solution to these limitations. In this work, we evaluate whether the computational cost of interval analysis yields significant enough improvements in BMC's performance to justify its use. In more detail, we quantify the benefits of interval analysis on two benchmarks: the Intel Core Power Management firmware and $9537$ programs in the ReachSafety category of the International Competition on Software Verification. Our results show that interval analysis is essential in solving $203$ unique benchmarks.
\end{abstract}
%

\section{Introduction}
\label{sec:introduction}

Software bugs can cause issues ranging from small nuisances to software exploits~\cite{microsoft-cpp}. For safety-critical systems, these bugs can cause high-cost damages. As an example, Southwest Airlines had a cost of over one billion dollars due to a cascade of bugs~\cite{SouthwerNetworkFail}. In a security report, the White House advocated for using formal methods to ensure software quality, emphasizing that testing is not enough~\cite{white-house:2024}.

Model checking~\cite{clarke1997model} is a formal methods technique that systematically explores a model against a safety property, ensuring the system's safety. In software analysis, bounded model checking (BMC)~\cite{clarke2001bounded} has been popularized in the industry by companies such as Amazon Web Services~\cite{cook2018model} and ARM~\cite{wu2024verifying}. The use of BMC has aided developers in finding bugs in their software. For instance, ESBMC found a bug in the Ethereum spec\footnote{\url{https://github.com/ethereum/consensus-specs/pull/3600}}. Additionally, BMC tools can produce safety proofs, i.e., giving further evidence that software can not fail the requirements for any input~\cite{white-house:2024}.

However, BMC has two major limitations: state-space explosion and lack of completeness~\cite{clarke2001bounded}. On the one hand, although BMC explores bounded traces, its memory complexity is still exponential in the length of the trace. On the other hand, proving program safety may require exploring an infinite number of traces. Modern BMC tools address these limitations using state-of-the-art SAT/SMT solvers and advanced proof strategies such as \textit{k}-induction~\cite{ramalho2019scalable}. At the same time, these solutions fall short on many real-world applications.

Interval analysis is a widely used technique in software verification to prove the \textit{absence} of vulnerabilities~\cite{cousot2021principles}. In this respect, it can further mitigate the limitations of BMC by removing unreachable program paths and showing that some assertions always hold. Furthermore, interval analysis can produce stronger program invariants, enabling most proof strategies to reach a verdict~\cite{ramalho2019scalable}. Unfortunately, computing precise intervals requires additional computational resources, which may nullify their advantages.

In this light, the problem of efficiently applying interval analysis to state-of-the-art BMC tools is still open. In this paper, we explore the inherent trade-off between these two techniques from three perspectives:
\begin{itemize}
    \item \textit{Application.} We determine the BMC stage (i.e., proof strategy, decision procedure) at which interval analysis introduces the best improvement.
    \item \textit{Precision.} We quantify the impact of computing tighter intervals on the overall verification results.
    \item \textit{Representation.} We compare two interval representations of machine integers: \textit{wrapped intervals} and \textit{integer lattices}.    
\end{itemize}

In more detail, we make the following contributions. First, we implement interval analysis inside ESBMC~\cite{menezes2024esbmc}, a state-of-the-art software model checker. Second, we measure the impact of different interval analysis techniques on BMC software verification. Third, we quantify the benefits of instrumenting the program with interval invariants and their effect on SMT-solving time and memory. Fourth, we evaluate the combination of interval analysis and BMC over a set of $9537$ C programs from the SV-COMP'23 \textit{ReachSafety} category. Finally, we demonstrate the improvements interval analysis brings when verifying the Intel Core Power Management firmware.

The remainder of this work is organized as follows: Section~\ref{sec:background} introduces the required background; Section~\ref{sec:goto_intervals} presents how to compute intervals through abstract interpretation; Section~\ref{sec:using-intervals-esbmc} describes how to apply the intervals inside a BMC framework; Section~\ref{sec:experiments} compares the experimental results of different interval approaches; Section~\ref{sec:conclusions} concludes and sketches a plan for future work.

\section{Background}
\label{sec:background}

\subsection{Bounded Model Checking (BMC)}
\label{sec:background_bmc}

BMC~\cite{biere1999symbolic} is a verification technique that symbolically explores a system against an LTL property~\cite{fisher2011introduction} producing satisfiability formulas. Although applications of this technique have initially focused on \emph{hardware}~\cite{clarke2018handbook}, it has found success in the \emph{software} domain~\cite{kroening2014cbmc,cordeiro2011smt}. The advantage of BMC is the limiting factor (i.e., bound) that helps with the state-explosion problem~\cite{biere1999symbolic}. The downside of BMC is that it is incomplete, as it cannot build a proof over an unbounded trace.

Briefly, BMC can be described as follows: Let \( \mathcal{P} \) be a program under verification, defined as a finite state transition system \( \mathcal{ST}=\left(S, R, I, T\right) \), where \( S \) represents the set of states, \( R \subseteq S \times S \) is the set of transitions between the states, \( I(s_i) \) defines a set of initial states \( s_i \in S \), and $T(s_i,s_{i+1})$ defines a set of pre-conditions (i.e., logical formulas) for reaching $s_{i+1}$ from $s_i$. In this context, a state \( s \in S \) consists of the program counter (\textit{pc}) and variable values. Given a safety property $\phi$, the BMC problem, \( B_{\phi}(k) \), determines whether there is a counterexample (i.e., a valid sequence of states \( \langle s_1, \ldots, s_k \rangle \)) up to length \( k \) that violates \( \phi \), i.e.:

\begin{equation*}\label{eq:bmc}
 B_{\phi}(k) = I(s_1) \wedge \bigwedge^{k-1}_{i=1} T(s_i, s_{i+1}) \wedge
\bigvee^{k}_{i=1} \neg \phi(s_i),
\end{equation*}

\noindent where the formula \( I(s_1)\wedge\bigwedge^{k-1}_{i=1} T(s_i, s_{i+1}) \) describes the execution of \( \mathcal{ST} \) up to \( k \) steps, while $\bigvee^{k}_{i=1} \neg \phi(s_i)$ checks whether the safety property $\phi$ is violated in one of the reachable states $s_i$.

When proving software safety for any $k$, BMC can still be applied by using unwinding assertions \cite{cordeiro2011smt} or a $k$-induction transformation  \cite{gadelha2017handling}:
\begin{itemize}
    \item \textit{Unwinding assertions.} This property checks whether $k$ is large enough to fully explore all program loops. That is, the negation of the loop conditions becomes assertions after the loop. If there is no counterexample, then all loops are fully explored (and the program is safe).
    \item \textit{$k$-induction.} All loop variables are \emph{havocked} (i.e., they are assigned new symbolic values), and then the property is checked against these unrestricted values. If there is no counterexample for the formula, then the property holds for any $k$. The $k$-induction main limitation is that the havoc procedure yields very weak invariants. 
\end{itemize}

\subsection{Interval Analysis}

Interval analysis consists of computing the minimal and maximal values for all variables in a program~\cite{cousot2021principles} such that $\forall x \in V . x \in  [\ubar{x},\bar{x}]$, where $V$ is the set of all program variables. The technique was introduced for scientific computing by Ramon E. Moore~\cite{moore1966interval,moore1979methods} to ensure bounds for rounding errors when computing floating-point values~\cite{kahan1996ieee}. 

\begin{figure}[t]
\begin{tcolorbox}[sidebyside]

\centering
Basic program \\

\begin{tikzpicture}[auto,
  node distance = 4mm,
  start chain = going below,
  every edge quotes/.append style = {auto, font=\footnotesize, inner sep=2pt},
  box/.style = {draw,rounded corners,blur shadow,fill=white,
        on chain,align=center}]
 \node[box] (b1)    {$x\leftarrow0$};      
 \node[box] (b2)    {$(x<100)$?};      
 \node[box] (b3)    {$x\leftarrow x+1$};  
 \node[box] (b4)    {assert($x > 50$)};     
 \begin{scope}[rounded corners,-latex]
  \path (b2.east) edge["F"][bend left=90] (b4.east)
  (b1) edge (b2) (b2) edge["T"] (b3);
  \draw (b3.230) -- ++(0,-0.3) -| ([xshift=-5mm]b2.west) |-
  ([yshift=3mm]b2.130) -- (b2.130);
 \end{scope}
\end{tikzpicture}  

\tcblower

\centering
Program intervals \\
\begin{tikzpicture}[auto,
  node distance = 4mm,
  start chain = going below,
  every edge quotes/.append style = {auto, font=\footnotesize, inner sep=2pt},
  box/.style = {draw,rounded corners,blur shadow,fill=white,
        on chain,align=center}]
 \node[box] (b1)    {$x : [0, 0]$};      
 \node[box] (b2)    {$x : [0, 100]$};      
 \node[box] (b3)    {$x : [1, 100]$};  
 \node[box] (b4)    {$x : [100, 100]$};     
 \begin{scope}[rounded corners,-latex]
  \path (b2.east) edge["F"][bend left=90] (b4.east)
  (b1) edge (b2) (b2) edge["T"] (b3);
  \draw (b3.230) -- ++(0,-0.3) -| ([xshift=-5mm]b2.west) |-
  ([yshift=3mm]b2.130) -- (b2.130);
 \end{scope}
\end{tikzpicture}

\end{tcolorbox}
\caption{Interval analysis of a program. On the left: a program with operations over the variable \emph{x}, and on the right: the computed intervals.}
\label{fig:interval-analysis-example}
\end{figure}
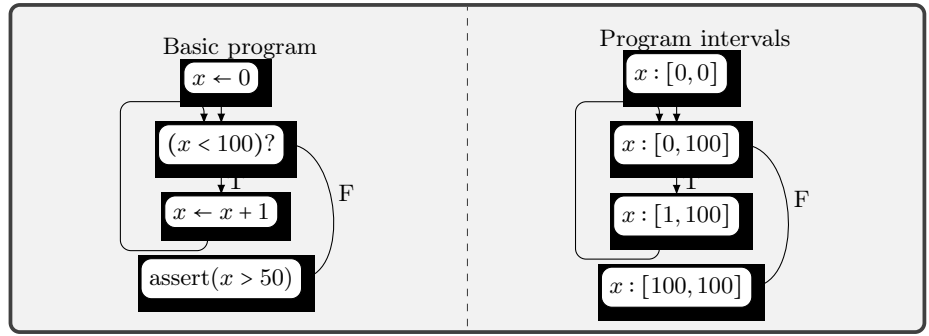

Figure~\ref{fig:interval-analysis-example} shows an example of an annotated program with intervals for all its statements. The program contains an assertion that $x > 50$. Looking at the intervals of the program, we can see that $x \in [100,100]$. Therefore, we can guarantee that the assertion always holds. If, for a statement, the computed interval is empty, then we can assume that this statement is unreachable~\cite{cousot2021principles}.

We can compute intervals in different ways: statically~\cite{cousot2021principles}, dynamically~\cite{cousot2021principles} and symbolically~\cite{sankaranarayanan2007program}. In this work, we focus on computing intervals statically through abstract interpretation.

\subsection{Computing Intervals through Abstract Interpretation}
\label{sec:background_abstract}

Abstract Interpretation is a static analysis framework to compute fixed-point lattices over systems. Proposed first by Cousot in (1977)~\cite{cousot:1977:abstract}, it can be applied to interval analysis by using specialized abstract domains~\cite{cousot:1977:abstract}. Within the same domain, different levels of precision can be applied to speed-up the procedure. For example, interval arithmetics~\cite{moore1979methods} and modular arithmetic for bit-vector semantics~\cite{muller2007analysis} may increase precision, while widening and other domain-specific techniques can accelerate the computation of fixed-point~\cite{blanchet2003static,singh2017fast,isaacson2012analysis,muller2007analysis}.

\subsubsection{Integer Domain ($\mathbb{I}$)} 
This is the classic domain over the Integers set~\cite{cousot2021principles}. Variables are assumed to range from $-\infty$ up to $+\infty$. This has the advantage of being straightforward when using arithmetic interval operators. However, it is not trivial to do bit-precise machine operations, e.g., sign typecasts and bitwise operations. An interval for a variable \emph{x} is represented as $x \in [l,u]$ where $l <= x <= u$. Figure~\ref{fig:integer-lattice} shows the complete lattice of this domain.

\subsubsection{Wrapped Domain ($\mathbb{W}$)} 
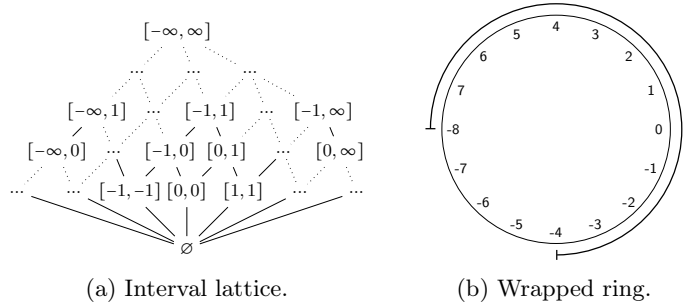
\begin{figure}[t]
\centering

\begin{subfigure}{.4\textwidth}
\resizebox{\columnwidth}{!}{%
\begin{tikzpicture}

  \node (top) at (0,0) {$[-\infty, \infty]$};

  \node [below left  of=top] (level1i0)  {...};
  \node [right of=level1i0] (level1i1)  {...};
  \node [right of=level1i1] (level1i2) {...};

  \draw [dotted] (top) -- (level1i0);
  \draw [dotted] (top) -- (level1i1);
  \draw [dotted] (top) -- (level1i2);


  \node [below left  of=level1i0] (level2i0)  {$[-\infty, 1]$};
  \node [right of=level2i0] (level2i1)  {...};
  \node [right of=level2i1] (level2i2)  {$[-1,1]$};
  \node [right of=level2i2] (level2i3)  {...};
  \node [right of=level2i3] (level2i4)  {$[-1, \infty]$};

  \draw [dotted] (level1i0) -- (level2i0);
  \draw [dotted] (level1i0) -- (level2i1);
  \draw [dotted] (level1i1) -- (level2i1);
  \draw [dotted] (level1i1) -- (level2i2);
  \draw [dotted] (level1i2) -- (level2i2);
  \draw [dotted] (level1i2) -- (level2i3);
  \draw [dotted] (level1i2) -- (level2i4);


  \node [below left  of=level2i0] (level3i0)  {$[-\infty, 0]$}; 
  \node [right of=level3i0] (level3i1)  {...};
  \node [right of=level3i1] (level3i2)  {$[-1,0]$};
  \node [right of=level3i2] (level3i3)  {$[0,1]$};
  \node [right of=level3i3] (level3i4)  {...};
  \node [right of=level3i4] (level3i5)  {$[0, \infty]$};

  \draw (level2i0) -- (level3i0);
  \draw [dotted] (level2i0) -- (level3i1);
  \draw [dotted] (level2i1) -- (level3i1);
  \draw [dotted] (level2i1) -- (level3i2);
  \draw (level2i2) -- (level3i2);
  \draw (level2i2) -- (level3i3);
  \draw [dotted] (level2i3) -- (level3i3);
  \draw [dotted] (level2i3) -- (level3i4);
  \draw [dotted] (level2i4) -- (level3i4);
  \draw (level2i4) -- (level3i5);

  \node [below left  of=level3i0] (level4i0)  {...}; 
  \node [right of=level4i0] (level4i1)  {...}; 
  \node [right of=level4i1] (level4i2)  {$[-1,-1]$};
  \node [right of=level4i2] (level4i3)  {$[0,0]$};
  \node [right of=level4i3] (level4i4)  {$[1,1]$};
  \node [right of=level4i4] (level4i5)  {...};
  \node [right of=level4i5] (level4i6)  {...};

  \draw [dotted] (level3i0) -- (level4i0);
  \draw [dotted] (level3i0) -- (level4i1);
  \draw [dotted] (level3i1) -- (level4i1);
  \draw (level3i1) -- (level4i2);
  \draw (level3i2) -- (level4i2);
  \draw (level3i2) -- (level4i3);
  \draw (level3i3) -- (level4i3);
  \draw (level3i3) -- (level4i4);
  \draw (level3i4) -- (level4i4);
  \draw [dotted] (level3i4) -- (level4i5);
  \draw [dotted] (level3i5) -- (level4i5);
  \draw [dotted] (level3i5) -- (level4i6);

  \node [below of=level4i3] (bottom)  {$\varnothing $}; 

  \draw (level4i0) -- (bottom);
  \draw (level4i1) -- (bottom);
  \draw (level4i2) -- (bottom);
  \draw (level4i3) -- (bottom);
  \draw (level4i4) -- (bottom);
  \draw (level4i5) -- (bottom);
  \draw (level4i6) -- (bottom);
\end{tikzpicture}%
}
\caption{Interval lattice.}\label{fig:integer-lattice}
\end{subfigure}%
\begin{subfigure}{.4\textwidth}
\begin{center}
\resizebox{0.7\columnwidth}{!}{%
\begin{tikzpicture}[
     auto,                
     node distance = 0cm, 
     bin/.style    = {rectangle, fill=white, text=black},
     dec/.style    = {draw=none, text=black},
    circ/.style    = {circle, red, thick, minimum size=5.25cm}
  ]

  \node[circ] (center) at (0,0) {};
  \draw (0,0) circle (2.5cm);

  \


  \foreach \angle / \dez in {%
    0/0, 22.5/1, 45/2, 67.5/3, 90/4, 112.5/5, 135/6, 157.5/7, 180/-8,
    202.5/-7, 225/-6, 247.5/-5, 270/-4, 292.5/-3, 315/-2, 337.5/-1}
    \draw (\angle:2.25cm) node [dec, font=\sffamily] {\dez};

    \draw[|-|, thick] (0,-2.75) arc (-90:180:2.75cm);
\end{tikzpicture}%
}%
\caption{Wrapped ring.}\label{fig:wrapped-lattice}
\end{center}
\end{subfigure}%
\caption{Abstract Domains.}
\label{fig:abstract-domains}
\end{figure}

In this domain, first introduced in~\cite{gange2015interval}, variables are restricted to a finite range, from the minimum representative for the type to its maximum. Wrapped interval sets are structured as a ring (see Figure~\ref{fig:wrapped-lattice}). 
An interval for a variable \emph{x} is represented as $x \in \langle l,u \rangle$ where $l <= x <= u$ iff $l <= u$, else $x >= l \vee x <= u$. Figure~\ref{fig:wrapped-lattice} contains an example of this domain for a 4-bit signed integer, the interval segment $\langle -4,-8 \rangle$ ranges from -4 up to all positive numbers and back to the negative values. 

\subsubsection{Boolean Domain} We adopt the C language convention, where boolean values are represented as $0$ for \textit{false} and $1$ for \textit{true}.\footnote{Note that in C, non-zero values are cast to \emph{true}, which has a numerical value of 1.} 
We use the same convention for boolean intervals, which yields the following three-valued logic: $[0,0]$ is \textit{false}, $[1,1]$ is \textit{true}, and $[0,1]$ is \textit{maybe}.

\section{Computing Intervals}
\label{sec:goto_intervals}

This section describes how to compute intervals formally. We first introduce the GOTO program language and then define an abstract interpreter for it.

\subsection{GOTO language}
\label{sec:goto-language}

We chose to base our work on a subset of the GOTO Language, which is an intermediate language used by the CBMC~\cite{kroening2014cbmc} and ESBMC~\cite{menezes2024esbmc} verifiers. The GOTO language has the following statements (see Appendix~\ref{app:algorithms} for the full grammar):
\begin{bnfgrammar}
statements \in $\textbb {S}$ : \textit{Program Statements} ;;
statements  ::= statements; statements : composition
| Assignment v expr : assign v to the result of expr
| Assumption bool-expr : assumes that bool-expr is true
| Assert bool-expr : asserts that bool-expr is true
| IfThenGoto bool-expr l : If bool-expr is true then goto l
| Label l : Sets jump location l
| Skip : Do nothing
\end{bnfgrammar}
Additionally, for simplicity in the explanations, we will assume that all variables are 32-bit signed integers (unsigned and other sizes are also supported).

\subsubsection{Program safety}

The result of any program execution is a truth value that represents whether any assertion in the program has failed. The procedure \texttt{eval} (see Algorithm \ref{alg:eval}) contains an implementation\footnote{The procedure is not specific to any verification technique. It is described for exemplification purposes.} to check the evaluation of a program execution where we assume that the \textit{eval-expr} procedure computes the values of expressions and the \textit{eval-state} procedure updates the environment (see Appendix~\ref{app:algorithms} for a definition of such semantics). Environments $\mathbb{Ev}: \text{Var} \rightarrow \mathbb{N}$ are maps that define values for all program variables for a given interpretation. For a program to be considered safe, the analyzer needs to prove that for the given program $s \in \mathbb{S}. \nexists \rho \in \mathbb{Ev}. \neg \text{eval}(\rho, s)$. In other words, an environment (interpretation) does not exist, leading to an assertion failure.

\begin{algorithm}%
\caption{Program evaluation}%
\begin{algorithmic}[1]%
\Procedure{eval}{$\rho \in \mathbb{Ev}, s \in \mathbb{S}$}      
    \State stmts $\leftarrow$ to-list(s)  \Comment{Sequentialize all statements in a list}
    \State labels $\leftarrow$ label-indexes(s)  \Comment{Generate a map for labels to stmts index}
    \State pc $\leftarrow$ 1 \Comment{Initialise Program Counter}
    \While{pc $\leq$ length(stmts)}
        \State s $\leftarrow$ stmts[s] \Comment{Get statement}
        \If{IsAssume(s)$\wedge $ \textbf{eval-expr}($\neg s.expr$, $\rho$)}
            \Return \textit{true}
        \EndIf   
        \If{IsAssert(s)$\wedge$ \textbf{eval-expr}($\neg s.expr$, $\rho$)}
            \Return \textit{false}
        \EndIf 
        \State $ \rho \leftarrow$ \textbf{eval-state}(S, $\rho$) \Comment{Update environment}

        \State pc $\leftarrow$ pc + 1 \Comment{Increment program counter}
        \If{IsIfThenGoto(s)$\wedge $ \textbf{eval-expr}($s.cond$, $\rho$)}
            \State pc $\leftarrow$ labels[s.label]
        \EndIf        
    \EndWhile
    \Return \textit{true} \Comment{Program finishes without any assertion fails}
\EndProcedure
\end{algorithmic}%
\label{alg:eval}%
\end{algorithm}%

\subsection{Abstract Domains for GOTO}
\label{sub:abstract-domain}

Now that we have a well-defined syntax and semantics for the GOTO language, we can adapt the classic abstract interpretation framework (see Section \ref{sec:background_abstract}) to it. This requires defining an abstract interpreter for the GOTO language. For this work, we require an abstract domain $\alpha$ that supports the following operations:
\begin{itemize}
    \item $\textit{Init}_\alpha()$. Initialize an interval for a state.
    \item $\textit{Transform}_\alpha(old, instruction)$. Change the current interval after interpreting the instruction.
    \item $\textit{Join}_\alpha(old,new)$. Merge the changes from previous intervals.
    \item $\textit{Widening}_\alpha(old,new)$. Widen the intervals based on the previous state.
\end{itemize}
In the remainder of this section, we will describe how these operations are defined for the Integer ($\mathbb{I}$) and Wrapped ($\mathbb{W}$) interval domains. 

\subsubsection{Initializing the Domain ($\textbf{Init}_\alpha$)}

When initializing an interval, we assume that the variable can take any representable value, i.e., the \emph{supremum} of the lattice. The supremum is unique for the $\mathbb{I}$ domain, as all variables are initialized as $(-\infty, +\infty)$. However, the $\mathbb{W}$ domain is not a lattice; therefore, we pick an arbitrary interval containing all possible representable values $\langle 0, -1 \rangle$.

\subsubsection{Transforming the Domain ($\textbf{Transform}_\alpha$)}
\label{sec:transform}

Program instructions can manipulate the domain variables with \emph{operations} (e.g. $x = y + 1$) or with \emph{restrictions} (e.g., $\textbf{if } x < 10 \textbf{ goto } 3$). Considering an abstract domain $\alpha$, we map variables into intervals by using the notion of abstract environments ($\mathbb{Ev}_\alpha : \text{Var} \rightarrow \alpha$). A program instruction can then update (\emph{transform}) this abstract environment.

\emph{Operations} are evaluations of program expressions, including arithmetic, bitwise, and casting (signed and unsigned). The interpretation of an expression $e \in \mathbb{E}$ over an abstract environment $\rho_\alpha \in \mathbb{Ev}_\alpha$ is defined as $\mathcal{E}_\alpha \llbracket e \rrbracket \rho_\alpha$. The $\mathbb{W}$ domain defines precise semantics for how these operations work, assuming that \emph{overflows} have modulo semantics. In contrast, the $\mathbb{I}$ domain defines arithmetic operations only.\footnote{Details of our implementation of integer interval arithmetics with bitwise operators are in Appendix \ref{app:warren}.} In practice, we may forfeit the precise semantics of these operations (i.e., using the $\textbf{Init}_\alpha$ function) to trade off precision for quicker computation (more on this in the next section). 

\emph{Restrictions} are conditions that are required to be met by the target instruction (such as the aforementioned \textbf{if} instruction destination). The result of a restriction is a sound approximation of the values that meet the condition. The restriction function over an abstract environment can be defined as $f_\alpha: \mathbb{E} \rightarrow \mathbb{Ev}_\alpha \rightarrow \mathbb{Ev}_\alpha$. This restriction can be implemented through the use of \textit{contractors}~\cite{aldughaim2020incremental,menezes2024esbmc} (for $\mathbb{I}$ domain) or by relying on the meet operator $\sqcap$ (or $\leq$) to restrict the intervals (e.g. $x \leq 10 \rightarrow [x_l, x_u] \sqcap [-\infty, 10] $), as both $\mathbb{I}$ and $\mathbb{W}$ contains a definition for it.

Finally, we can define how an instruction alters the abstract state through the $\mathcal{S}_\alpha$ operator. Let $\mathcal{S}_\alpha \llbracket S \rrbracket \rho_\alpha$ be the interpretation of a statement $S \in \mathbb{S}$ within an abstract environment $\rho_\alpha \in \mathbb{Ev}_\alpha$,  defined as:

\begin{align*}
    \mathcal{S_\alpha} \llbracket s_1;s_2 \rrbracket \rho_\alpha &\triangleq \mathcal{S_\alpha} \llbracket s_2 \rrbracket(\mathcal{S}_\alpha \llbracket s_1 \rrbracket \rho_\alpha)  \\
    \mathcal{S_\alpha} \llbracket \text{Assignment v e} \rrbracket \rho_\alpha &\triangleq   \forall x \in \mathbb{V}.  x = v \rightarrow \mathcal{E}_\alpha \llbracket e \rrbracket \rho_\alpha \wedge x \neq v \rightarrow \rho_\alpha(x) \\
    \mathcal{S_\alpha} \llbracket \text{Assume e} \rrbracket \rho_\alpha &\triangleq f_\alpha(e, \rho_\alpha) \\
    \mathcal{S_\alpha} \llbracket \text{Assert e} \rrbracket \rho_\alpha &\triangleq f_\alpha(e, \rho_\alpha) \\
    \mathcal{S_\alpha} \llbracket s \in \mathbb{S} \rrbracket \rho_\alpha &\triangleq \rho_\alpha
\end{align*}%

\subsubsection{Joining Intervals ($\textbf{Join}_\alpha$)}

After transforming an interval, we require a function to merge intervals from different program paths. Before defining the join operation, we need to define the ordering operator represented as $i_p \sqsubseteq i_n$, which is true \textit{iff} $i_p$ is at a higher level than $i_n$ in the domain lattice. Intuitively, the ordering operator represents whether interval $i_p$ is a sound abstraction of interval $i_n$. The join operation over two intervals $i_0$ and $i_1$ consists of computing an interval $i_j = i_0 \sqcup i_1$ such that $i_j \sqsubseteq i_0 \wedge i_j \sqsubseteq i_1$. Both $\mathbb{I}$ and $\mathbb{W}$ contain descriptions for such operators.

\begin{algorithm}[t]
\caption{Computation of the abstract domain}
\begin{algorithmic}[1]
\Procedure{computeAbs}{$s \in \mathbb{S}$}      
    \State stmts $\leftarrow$ to-list(s);  labels $\leftarrow$ label-indexes(s) ; work-list $\leftarrow \{1\}$; $P_\alpha \leftarrow \emptyset$ 
   
    \While{$\neg$ work-list.empty()}
        \State pc $\leftarrow$ work-list.pop() ; s $\leftarrow$ stmts[pc]     

        \State state $\leftarrow \neg P_\alpha$.contains(pc) ? $\textbf{Init}_\alpha$ : domain-stmt[pc]
        \State next $\leftarrow \{ \text{pc} + 1 \}$
        \If{IsIfThenGoto(s)} \State next.add(labels[s.label])
        \EndIf 
        \While{$\neg$ next.empty()}
            \State to $\leftarrow$ next.pop()
            \State new-state  $\leftarrow \textbf{Transform}_\alpha(\text{old-state}, \text{to})$  
            \State old-state $\leftarrow$ $P_\alpha$[to]
            \If {$\exists x \in \text{old-state } . x = \bot \vee \exists y \in \text{new-state } . y = \bot$}
                \State domain-stmt[to] $\leftarrow \text{new-state}$
            \Else
                \State domain-stmt[to] $\leftarrow  \forall x \in \text{new-state } .\textbf{ Join}_\alpha(\text{old-state}(x),\text{new-state}(x))$
            \EndIf
            \If{old-state $\neq$ new-state}
                    \State \textbf{Widening}$_\alpha$(old-state, new-state) ; work-list.add(\text{to})
        \EndIf
        \EndWhile
    \EndWhile
    \Return $P_\alpha$
\EndProcedure%
\end{algorithmic}%
\label{alg:compute_abs}
\end{algorithm}%

\subsubsection{Widening Intervals ($\textbf{Widening}_\alpha$)}

The widening operator consists of heuristics that accelerate reaching the fixed-point. 
For the integer domain, the widening consists of rapidly climbing the lattice towards the supremum. The formal definition of such operation is defined as:

\[ \text{extrapolate}([l_p, u_p], [l_n, u_n]) = \begin{cases} 
      (-\infty, +\infty) & l_n < l_p \wedge u_n > u_p \\
      (l_n, +\infty) & u_n > u_p \\
      (-\infty, u_n) & l_n < l_p  \\
      [l_n, u_n] & \textit{otherwise}
   \end{cases}
\]

 For the wrapped domain, the widening consists of expanding the ring on both sides (by a factor of $2$) until the ring is fully completed. To further understand how the ring is expanded, we refer to the original work~\cite{gange2015interval}.

\subsection{Abstract Interpreter for GOTO}
\label{sec:abstract-interpreter}
 
Now that we have described how the domains behave on GOTO statements, we can finally define the associated abstract interpretation procedure. More specifically, the procedure computes an abstract environment (see Section \ref{sec:transform}) over all program statements resulting in the map $P_\alpha : \mathbb{S} \rightarrow \mathbb{Ev}_\alpha$. 

We present the pseudocode of our abstract interpreter in Algorithm \ref{alg:compute_abs}. The procedure \texttt{computeAbs} starts from an empty domain and a work list containing the first statement (Line 2). Whenever we reach a statement we have never visited, we initialize its state with the $\textbf{Init}_\alpha$ procedure (Line 5). Then, the abstract interpreter processes any domain transformations (Line 11). The rules for merging conflicting states are in Lines 12 to 16. When the merged state differs from the previous one, the interpreter adds the next instruction to the work list (line 18). We have reached a fixed-point when the work list is empty (Line 3).

\subsubsection{Storing the Domain} 
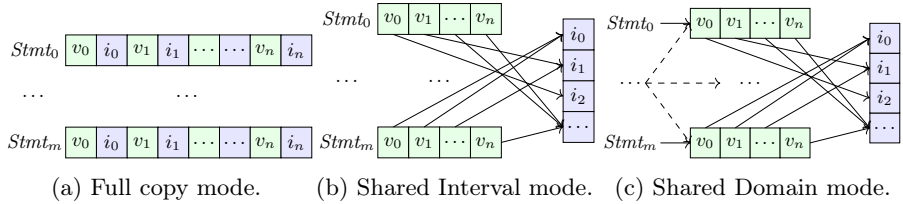
\begin{figure}[t]
\centering
\begin{subfigure}{.34\textwidth}
    \centering
    \resizebox{\columnwidth}{!}{%
    \begin{tikzpicture}[scale=0.5]
    \foreach \x in {0,...,3} {
        \draw[fill=green!10](2*\x, 0) rectangle  ++(1,1);
        \draw[fill=blue!10](2*\x+1, 0) rectangle  ++(1,1);
    }
    \foreach \x in {0,...,1} {
        \node at (2*\x + 0.5, 0.5) { $v_\x$};
        \node at (2*\x + 1.5, 0.5) { $i_\x$};
    }
    \node at (2*2 + 0.5, 0.5) { $\ldots$};
    \node at (2*2 + 1.5, 0.5) { $\ldots$};
    \node at (2*3 + 0.5, 0.5) { $v_n$};
    \node at (2*3 + 1.5, 0.5) { $i_n$};
    \node at (-1,0.5) { $\textit{Stmt}_0$};

    \node at (-1,-1) { $\dots$};
    \node at (4,-1) { $\dots$};

     \foreach \x in {0,...,3} {
        \draw[fill=green!10](2*\x, -3) rectangle  ++(1,1);
        \draw[fill=blue!10](2*\x+1, -3) rectangle  ++(1,1);
    }

    \foreach \x in {0,...,1} {
        \node at (2*\x + 0.5, -2.5) { $v_\x$};
        \node at (2*\x + 1.5, -2.5) { $i_\x$};
    }

    \node at (2*2 + 0.5, -2.5) { $\ldots$};
    \node at (2*2 + 1.5, -2.5) { $\ldots$};
    \node at (2*3 + 0.5, -2.5) { $v_n$};
    \node at (2*3 + 1.5, -2.5) { $i_n$};
    \node at (-1,-2.5) { $\textit{Stmt}_m$};  
  \end{tikzpicture}%
    }%
    \caption{Full copy mode.}
    \label{fig:full-domain}
\end{subfigure}%
\begin{subfigure}{.31\textwidth}
    \centering
    \resizebox{\columnwidth}{!}{%
    \begin{tikzpicture}[scale=0.5]
    \foreach \y in {0,...,3} {
        \draw[fill=blue!10](0, \y) rectangle  ++(1,-1);        
    }
    \node at (0.5,-0.5) { $\ldots$};
    \node at (0.5,0.5) { $i_2$};
    \node at (0.5,1.5) { $i_1$};
    \node at (0.5,2.5) { $i_0$};

    \foreach \x in {0,...,3} {
        \draw[fill=green!10](\x - 5, 2.5) rectangle  ++(-1,1);        
    }
    \node at (-5.5,3) { $v_0$};
    \node at (-4.5,3) { $v_1$};
    \node at (-3.5,3) { $\ldots$};
    \node at (-2.5,3) { $v_n$};

    \node at (-4,1) { $\ldots$};

    \foreach \x in {0,...,3} {
        \draw[fill=green!10](\x - 5, -1.5) rectangle  ++(-1,1);        
    }
    \node at (-5.5,-1) { $v_0$};
    \node at (-4.5,-1) { $v_1$};
    \node at (-3.5,-1) { $\ldots$};
    \node at (-2.5,-1) { $v_n$};

    \draw[->]  (-5.5, 2.5) -- (0,0.5);
    \draw[->]  (-4.5, 2.5) -- (0,1.5);
    \draw[->]  (-3.5, 2.5) -- (0,-0.5);
    \draw[->]  (-2.5, 2.5) -- (0,-0.5);

    \draw[->]  (-5.5, -0.5) -- (0,2.5);
    \draw[->] (-4.5, -0.5) -- (0,2.5);
    \draw[->]  (-3.5, -0.5) -- (0,1.5);
    \draw[->]  (-2, -1) -- (0,-0.5);

    \node at (-7,3) { $\textit{Stmt}_0$};
    \node at (-7,1) { $\ldots$};
    \node at (-7,-1) { $\textit{Stmt}_m$};    
  \end{tikzpicture}%
    }%
   
    \caption{Shared Interval mode.}
    \label{fig:shared-interval}
\end{subfigure}%
\begin{subfigure}{.34\textwidth}
    \centering
    \resizebox{\columnwidth}{!}{%
    \begin{tikzpicture}[scale=0.5]
    \foreach \y in {0,...,3} {
        \draw[fill=blue!10](0, \y) rectangle  ++(1,-1);        
    }
    \node at (0.5,-0.5) { $\ldots$};
    \node at (0.5,0.5) { $i_2$};
    \node at (0.5,1.5) { $i_1$};
    \node at (0.5,2.5) { $i_0$};

    \foreach \x in {0,...,3} {
        \draw[fill=green!10](\x - 5, 2.5) rectangle  ++(-1,1);        
    }
    \node at (-5.5,3) { $v_0$};
    \node at (-4.5,3) { $v_1$};
    \node at (-3.5,3) { $\ldots$};
    \node at (-2.5,3) { $v_n$};

    \node at (-4,1) { $\ldots$};

    \foreach \x in {0,...,3} {
        \draw[fill=green!10](\x - 5, -1.5) rectangle  ++(-1,1);        
    }
    \node at (-5.5,-1) { $v_0$};
    \node at (-4.5,-1) { $v_1$};
    \node at (-3.5,-1) { $\ldots$};
    \node at (-2.5,-1) { $v_n$};

    \draw[->]  (-5.5, 2.5) -- (0,0.5);
    \draw[->]  (-4.5, 2.5) -- (0,1.5);
    \draw[->]  (-3.5, 2.5) -- (0,-0.5);
    \draw[->]  (-2.5, 2.5) -- (0,-0.5);

    \draw[->]  (-5.5, -0.5) -- (0,2.5);
    \draw[->] (-4.5, -0.5) -- (0,2.5);
    \draw[->]  (-3.5, -0.5) -- (0,1.5);
    \draw[->]  (-2, -1) -- (0,-0.5);

    \node at (-8,3) { $\textit{Stmt}_0$};
    \node at (-8,1) { $\ldots$};
    \node at (-8,-1) { $\textit{Stmt}_m$};

    \draw[->]  (-7, 3) -- (-6,3);
    \draw[->]  (-7, -1) -- (-6,-1);
    \draw[->, dashed]  (-7.5, 1) -- (-6,3);
    \draw[->, dashed]  (-7.5, 1) -- (-5,1);
    \draw[->, dashed]  (-7.5, 1) -- (-6,-1);

  \end{tikzpicture}
    }%
    \caption{Shared Domain mode.}
    \label{fig:shared-domain}
\end{subfigure}%
\caption{Domain data structures. (a) a data structure that stores intervals for all variables and statements; (b) a data structure where the intervals are shared between all statements to avoid redundancy; (c) a data structure that improves over (b) by sharing groups of intervals.}
\label{fig:domains-memory}
\end{figure}

Executing the abstract interpretation in a memory-efficient way is not trivial.
We need to define a data structure that keeps track of the interval map $P_\alpha$ for all variables in each statement. A na\"ive implementation would lead to a $O(MN)$ memory complexity, where $M$ is the number of statements and $N$ is the number of variables (see Figure \ref{fig:full-domain}).
To circumvent this, we opt out for a lazy approach that only tracks the used domain references (see Figure \ref{fig:shared-interval}). We combine this approach with copy-on-write (COW) semantics to avoid unnecessary copies (see Figure \ref{fig:shared-domain}). 
Furthermore, we reduce the size of our data structure by only tracking the variables that are not \emph{supremum} and using a flag to record whether the domain contains an \emph{infimum}.

\section{Using Intervals in ESBMC}
\label{sec:using-intervals-esbmc}

To evaluate the techniques presented in Section~\ref{sec:goto_intervals}, we extend the Efficient SMT-based Bounded Model Checker (ESBMC), a state-of-the-art industrial-scale software verifier~\cite{menezes2024esbmc}. ESBMC provides the infrastructure to effectively implement interval analysis (i.e., $k$-Induction and SMT). We expect our methodology to easily apply to other state-of-the-art BMC tools.

\subsection{Architecture}

\begin{figure*}[t]
    \centering
    \resizebox{\columnwidth}{!}{%
    \begin{tikzpicture}[>=latex', very thick, align=center]
        \tikzstyle{every node}=[font=\Large]
        \tikzset{block/.style= {draw, rectangle, align=center,minimum width=4cm,minimum height=2cm, rounded corners=0.1cm},
        }
        \tikzstyle{decision} = [draw, diamond, aspect=2, fill=black!20]
        \tikzstyle{descriptionblock} = [draw, rectangle  split,  rectangle split parts=2, rectangle split part fill={black!20,black!10}, 
              minimum height=2 cm, text width=6.75cm]

        \tikzstyle{processblock} = [draw, rectangle, fill=black!20, minimum height=2 cm, text width=6.75cm]

        \node [coordinate]  (start) {One};
        \node[coordinate, below = 1cm of start] (last){last};
        \node [coordinate, right = 0.5cm of start] (ADL){ADL};
        \node [coordinate, right = 0.5cm of last] (BUL){BUL};
        \node [coordinate, above = 1.5cm of BUL] (AUL){AUL};
        \node [coordinate, below = 0.5cm of BUL] (BDL){BDL};
        \node [coordinate, below = 1.80cm of BDL] (BLL){BLL};

        \node [processblock,text width=3cm, left =\blockdistance of BDL] (Jimple) at (0,2) {\textbf{\textit{GOTO converter}}
        };
        
        \node [descriptionblock,text width=3cm, below =\blockdistance of Jimple] (Start) {\textbf{\textit{Input Program}}\nodepart{two}
        \vspace{-0.2cm}
        \begin{itemize}
            \item C
        \end{itemize}
        };
        
		\node [descriptionblock,text width= 6.5cm, right = 0.5cm of AUL] (step1) { \textbf{Abstract Interpretation} \nodepart{two}
		GOTO program optimization
		\begin{enumerate}[noitemsep]
		    \item Interval Analysis computation;
		    \item Interval optimizations;
		    \item Interval instrumentations.
		\end{enumerate}
		};
		
		\node [descriptionblock,text width= 8cm ,right = \blockdistance of step1] (step2) {\textbf{Symbolic Execution}\nodepart{two}
		Executes the GOTO program symbolically/runtime
		\begin{enumerate}[noitemsep]
		    \item Generates execution paths, including context switches.
                \item Add verification conditions.
		\end{enumerate}
		};
         
         \node [decision, right =\blockdistance of step2, node distance=7cm] (analyze) {\begin{tabular}{c}
          \textbf{Decision}\\\textbf{Procedure}
         \end{tabular}};
         
    	\node[block, fill=yellow!10, right = 1.25 cm of analyze] (unknown) {UNKNOWN};
    	
        \node[block, fill=green!10, above = 1 cm of unknown] (safe) {SAFE};
         
         \node[block, fill=red!10, below  = 1 cm of unknown] (unsafe){UNSAFE}; 
        
        \node [coordinate, right = 0.3cm of step2] (A2){One A2};
        \node[coordinate, below= 2 of A2] (B3){One B3};
        \node[coordinate, right = 7cm of B3](B4){B4};
        
        \path[draw, ->, line width=3pt]
            (Start) edge (Jimple)
            (Jimple) edge ++(3.3cm,0)
            
            (step1) edge (step2)
            (step2) edge (analyze)
            (analyze) edge (unknown)
            (analyze.north) -- ++(0,1.5cm) edge (safe.west)
            (analyze.south) -- ++(0,-1.5cm) -- (unsafe.west);
     \begin{pgfonlayer}{background}
		\path (step1.west |- step1.north)+(-0.75,0.8) node (a) {};
		\path (analyze.south -| analyze.east)+(0.25,-3.5) node (b) {};
		\end{pgfonlayer}

    \begin{pgfonlayer}{background}
		\path[fill=black!3,rounded corners, draw=black!50, dashed]
		(a) rectangle (b);
		\end{pgfonlayer}
		
        \node[draw] at (2.3,-4) {ESBMC};
    \end{tikzpicture}%
    }%
    \caption{The ESBMC architecture with interval analysis.} 
    \label{fig:ESBMC-architecture}
\end{figure*}
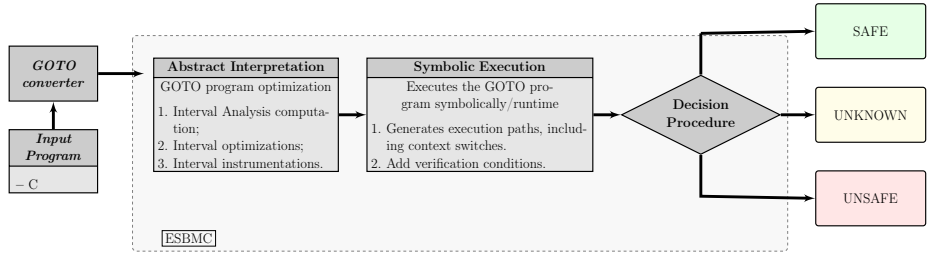

Figure~\ref{fig:ESBMC-architecture} shows the portion of ESBMC architecture that is relevant for our interval analysis extension (refer to \cite{menezes2024esbmc} for a full description). The flow can be described as follows. First, we begin with the source code of the input program, which is compiled into the GOTO language. Second, we compute intervals over the GOTO program using abstract interpretation. Third, we use the intervals to further transform the program. Fourth, we symbolically execute the transformed GOTO program, resulting in a formula. Finally, we evaluate whether the program is safe via a \textit{Decision Procedure}~\cite{kroening2016decision}. The \textit{Decision Procedure} will systematically increase the BMC bound $k$ until either a vulnerability is found, a correctness proof is found (by either the unwinding assertions or $k$-induction), or resource exhaustion.

In the remainder of this section, we explain how we transform the GOTO program with the information obtained by computing the intervals.

\subsection{Optimizations}
\label{sec:optimisations}

Once we have access to the intervals, we can optimize the size of the GOTO program by slicing and folding the code. Here, we cover two crucial optimizations: singleton propagation and dead code removal.

When evaluating an expression, the resulting interval may be 
a singleton, i.e., only one value lies within the interval. In such cases, we can replace the entire content of the expression with the
singleton value.
An example of this optimization is shown in Figure~\ref{lst:optimized}, where the assertion value is the singleton $[1,1]$. A side effect of singleton propagation is that it yields a tunable \emph{constant propagation} algorithm. The precision level of the interval analysis has a direct impact on the effectiveness of the propagation.

Finally, the intervals can be used to remove unreachable instructions. These can be easily detected by checking whether $P_a$ of the statement contains an \emph{infimum}. Converting such instructions to \emph{Skip} simplifies the program analysis by removing irrelevant branches. In our experimental analysis of Section \ref{sec:SV-COMP}, we identify a subset of benchmarks where this optimization can remove entire loops.

\subsection{Invariant Instrumentation}
\label{sec:invariant}

ESBMC relies on SMT solvers, which are responsible for proving the satisfiability of the formula. Introducing invariants generally restricts the search space that the solver needs to verify. At the same time, introducing too many invariants might dramatically increase the size of SMT formulae, thus causing the solver to be slower. To explore this trade-off, we prepare three different levels of instrumentation. We list them below in order of verbosity:

\begin{itemize}
    \item \textit{Loop Instructions.} We instrument each loop with the intervals of all variables appearing inside it by adding extra assumptions before and after it.
    \item \textit{Guard Instructions (full or local).} We instrument each guard instruction (assumption, assertion, condition) with an extra assumption, which covers all variables in the program (full) or just the variables in the guard (local).
    \item \textit{All Instructions (full or local).} We instrument every instruction in the program with an extra assumption, which covers all variables in the program (full) or just the variables in the corresponding statement (local).
\end{itemize}
Given the above, we can now apply interval analysis to program verification.

\begin{figure}[t]%
\begin{tcolorbox}[sidebyside]%
 
  \centering
  \begin{minted}{c}
void foo() {
  int a = * ? 4 : 6; // a: [4.6]
  assert(a + 2 >= 6);
}
  \end{minted}

\tcblower
  \centering
  \begin{minted}[escapeinside=||]{c}
void foo() {
  int a = * ? 4 : 6; // a: [4.6]
  assert(|\colorbox{green}{1}|);
}
\end{minted}
\end{tcolorbox}%
 \caption{Optimization example. The example contains the original program (left) and its optimized form (right).}%
 \label{lst:optimized}%
\end{figure}%

\section{Experimental Evaluation}
\label{sec:experiments}

In this work, we pose the question of how to use interval analysis inside a BMC framework and whether its additional processing cost results in better performance. 
More specifically, we evaluate the impact of interval analysis on two software verification benchmarks:
the benchmarks used by the Software Verification Competition (SV-COMP) in Reachability~\cite{beyer2024state}, and an industrial case of a power management firmware. This section is divided in three parts: Intel Power Management Firmware, SV-COMP reachability set, and a more in-depth analysis of affected benchmarks. 

\subsection{Intel Core Power Management Firmware}
\label{sec:firmware}

Intel routinely employs ESBMC to automate firmware analysis. In the past, ESBMC has been applied to the Authenticated Code Module \cite{futral2013fundamental}, where it found over $30$ vulnerabilities. ESBMC is part of the CI pipeline for developing microcode for the Core family of processors~\cite{gwennap1997p6}.

In the interest of expanding its use, Intel assessed the performance of ESBMC on the \textit{Core Power Manager}. This piece of software controls the CPU frequency to reduce thermal damage. More specifically, it has the following features:

\begin{itemize}
    \item An event-driven behavior where the application reacts to hardware triggers by executing floating-point computations.
    \item A harness that simulates the triggers with an infinite non-deterministic loop;
    \item Around 300 global variables that flag various hardware events;
    \item Around 190,000 lines of code split between 60 C files and 50 headers;
\end{itemize}

Without interval analysis, this benchmark is challenging for ESBMC and results in timeout after 3 days. However, adding interval analysis at the GOTO level allows ESBMC to reach a verdict in 8 hours. We report the specific combination of flags required to achieve this result in Appendix \ref{app:esbmc-options}. 
In the next section, we show that interval analysis improves the ESBMC performance across a wider variety of domains.



\subsection{SV-COMP Reachability Set}
\label{sec:SV-COMP}

The International Competition on Software Verification (SV-COMP), established in 2012, is one of the major driving forces of innovation in the software verification community~\cite{beyer2024state}. Over the years, the competition has collected many software verification benchmarks divided into run-sets. Each run-set is designed to evaluate state-of-the-art verifiers against a specific software vulnerability, e.g., memory corruption, arithmetic overflows, and assertion failures. 

\begin{table}[t]
	\caption{SV-COMP scoring systems without witness validation.
 } \label{table:score-system} 
			\begin{tabularx}{\textwidth}{|X|c|}
				\hline
				\textbf{Result} & \textbf{Score}  \\
				\hline 
                Correctly identifying that a benchmark has no vulnerabilities \textbf{(CT)}  & +2    \\ \hline  
                Correctly identifying that a benchmark has a vulnerability \textbf{(CF)}  & +1    \\ \hline  
                Mislabeling a vulnerable program as safe \textbf{(IT)} & -32    \\ \hline  
                Mislabeling a safe program as vulnerable \textbf{(IF)} & -16    \\ \hline
                The tool crashes due to errors or resource exhaustion & 0 \\ \hline
			\end{tabularx}
\end{table}

\subsubsection{Benchmark Description} For this work, we choose the benchmarks in the \emph{ReachSafety} run-set of SV-COMP'23. ReachSafety programs do not contain any undefined behavior according to the C99 standard, including overflows and memory safety issues. Furthermore, ReachSafety programs are the only ones compatible with the $k$-induction strategy, as implemented in ESBMC.

\subsubsection{Experimental Setup} We run our experiments in a similar fashion to SV-COMP~\cite{beyer2024state}. Specifically, we execute the benchmarks with version 3.21 of the \textit{benchexec} tool~\cite{beyer2019reliable}. We limit the computational resources to 120 seconds of CPU time, 1 CPU core and 6GB for memory. We also use a similar scoring system (see Table~\ref{table:score-system}), but we omit the witness validation phase as it would obscure our results. We execute all experiments on a KVM machine running Ubuntu 20.04 with Kernel 5.4.0-177-generic. Our hardware is a 32-core Intel(R) Xeon(R) CPU E5-2620 v4 @ 2.10GHz with 170GiB of RAM.


\subsubsection{Experimental Goals} Here, our goal is to confirm whether interval analysis helps ESBMC solve more benchmarks. To this end, we compare two settings:

\begin{itemize}
    \item \textit{Baseline.} This setting consists of ESBMC without any interval computation. See Appendix~\ref{app:esbmc-options} for an in-depth description of the specific flags.
    \item \textit{Intervals.} This setting enables the use of interval analysis. Specifically, we enable \textit{Integer Domain}, \textit{Optimizations} (see Section \ref{sec:optimisations}) and \textit{Loop Instructions} instrumentation (see Section \ref{sec:invariant}) on top of the \textit{Baseline} setting. In Section \ref{sec:invariant-mode}, we demonstrate why this configuration is optimal. 
\end{itemize}


\begin{table}[t]
	\caption{Aggregate results on SV-COMP reachability benchmarks. 
 We report the number of unique benchmarks solved by each ESBMC Setting in parenthesis.
 } \label{table:overall-results} 
			\begin{tabularx}{\textwidth}{|X|c|c|c|c|c|c|c|c|}
				\hline
				\textbf{Setting} & \textbf{CT}  & \textbf{CF } & \textbf{IT} & \textbf{IF} & \textbf{Timeout} & \textbf{Memory-out} & \textbf{Crashes} & \textbf{Score}  \\
				\hline 
                Baseline & 2611 (16) & 1657 (11) & 7 (0) & 0 (0) & 5112 (294) &  119 (4) & 31 (0) & 6655 \\ \hline
                Intervals  & 2789 (194) & 1654 (8) & 7 (0) & 0 (0) &  4848 (30) & 208 (93) & 31 (0) & 7008 \\ \hline
			\end{tabularx}
\end{table}

\begin{table}[t]
	\caption{Computational cost of interval analysis.
 We report mean and standard deviation on the SV-COMP benchmarks solved by both ESBMC Settings.
 } \label{table:overall-results-statistics} 
			\begin{tabularx}{\textwidth}{|X|c|c|c|c|}
				\hline
				\textbf{Setting} & \textbf{CPU (s)} & \textbf{Memory (MiB)} & \textbf{Preprocessing (s)} & \textbf{SMT Solving (s)} \\
				\hline 
                Baseline & $10.9\pm19.9$ & $116\pm166$ & $0.04\pm0.5$ & $6.15\pm15.4$ \\
                \hline
                Intervals & $11.1\pm20.1$ & $132\pm288$ & $0.5\pm3.4$ & $5.94\pm15.2$ \\
                \hline
			\end{tabularx}
\end{table}

\subsubsection{Overall Improvements} Overall, interval analysis improves the performance of ESBMC by 5\% (see Table \ref{table:overall-results}). 
The majority of the improvements come from the ability of interval analysis to prove the safety of additional benchmarks rather than finding violations in the unsafe ones.
The few incorrect results are unrelated to the interval analysis introduction in ESBMC. Indeed, the \textit{Baseline} version of ESBMC produces the same incorrect results.
We reported the issue to the developers, but no solution was implemented at the time of this writing.\footnote{\url{https://github.com/esbmc/esbmc/issues/1652}}

\subsubsection{Resource Consumption} Computing intervals increases the computational load (see Section \ref{sec:abstract-interpreter}). Here, we quantify the cost of interval analysis by comparing it with the baseline setting in ESBMC. In Table~\ref{table:overall-results-statistics}, we report the time and memory consumption on all benchmarks on which both versions of ESBMC reached the same verdict. 
From the data, we can see a slight increase in the CPU time and a moderate increase in memory consumption. 

\begin{figure}[t]%
\begin{center}%
\begin{tikzpicture}%
    \begin{axis}[
    	xlabel=Number of Unique Benchmarks Solved,
        xbar=1pt, 
        bar width=5pt, 
        y=14pt, 
        ytick=data,
        symbolic y coords={{ProductLines},{Loops},{Hardware},{ECA},{DeviceDrivers},{Combinations},{Arrays}},
        width=11cm, 
        height=5cm,
        ymajorgrids = true,
        xmin=0,
        xmax=90,
    ]
        \addplot[draw=matlab_color_1, fill=matlab_color_1!50] coordinates {(0,{ProductLines}) (0,{Loops}) (6,{Hardware}) (18,{ECA})  (1,{DeviceDrivers}) (0,{Combinations}) (1,{Arrays})};
        \addplot[draw=matlab_color_3, fill=matlab_color_3!50] coordinates {(89,{ProductLines}) (6,{Loops}) (1,{Hardware}) (30,{ECA}) (6,{DeviceDrivers}) (54,{Combinations}) (0,{Arrays})};
        \legend{Baseline,Intervals}
    \end{axis}%
\end{tikzpicture}%
\caption{Unique results per category. Y-axis consists of categories where the unique results were identified and X-axis is the quantity.} \label{fig:unique-per-category}%
\end{center}%
\end{figure}
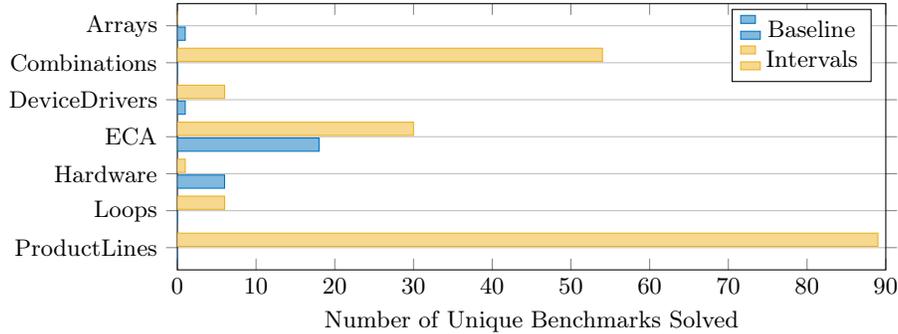%

\subsubsection{Unique Benchmarks Solved} In Figure \ref{fig:unique-per-category}, we isolate the subset of the SV-COMP benchmarks that are particularly affected by the introduction of interval analysis. 
There, we report the number of unique benchmarks that are only verifiable by either the \textit{Baseline} (no intervals) or the \textit{Interval} setting, but not both. More in detail, we observe unique results in the following sub-categories:
\begin{itemize}
    \item \textit{Combinations, DeviceDrivers, ECA, ProductLines.} These benchmarks contain manipulation of (global) ``guard'' variables in an event system, i.e., variables that represent that an event has happened. 
    Computing intervals over these variables reduces the number of paths BMC has to reason on.
    \item \textit{Loops.} These benchmarks contain infinite loops. Interval analysis introduces useful invariants for the \emph{k-induction} algorithm, while optimization removes loops that do not affect the relevant assertions in the program. 
    \item \textit{Hardware.} These benchmarks are generated from formal hardware descriptions and contain infinite loops with thousands of variables. As such, computing intervals result in memory and time exhaustion. 
    \item \textit{Arrays.} These benchmarks are solved within 5s from the timeout, thus causing a small number of spurious unique results. 
\end{itemize}
We use this reduced set of benchmarks for our detailed analysis in Section \ref{sec:invariant-mode}.


\subsection{Comparing Intervals Modes}
\label{sec:invariant-mode}

Sections \ref{sec:SV-COMP} and \ref{sec:firmware} show positive results for interval analysis. However, we obtained them under very different experimental setups (see Appendix \ref{app:esbmc-options}). Thus, identifying the best way to use intervals inside a BMC framework remains open. Here, we answer it by running additional experiments on the SV-COMP sub-categories of Figure \ref{fig:unique-per-category}. We structure our experiments according to the three perspectives of interval application, precision, and representation.



\subsubsection{Application} There are two main strategies when using the intervals: optimizations (Section \ref{sec:optimisations}) and instrumentation (Section \ref{sec:invariant}). In Table~\ref{table:application-results}, we demonstrate their separate impact on the performance of ESBMC.
The results show that:
\begin{itemize}
    \item None of the settings helps ESBMC identify additional safety violations (CF).
    \item Optimization has similar results to instrumentation. This shows that these two strategies have considerable overlap. 
    \item The Loop Instructions setting achieves the best results. We use this setting throughout Section~\ref{sec:SV-COMP}. 
\end{itemize}

\begin{table}[t]
	\caption{Comparison between optimization and instrumentation. 
 } \label{table:application-results} 
			\begin{tabularx}{\textwidth}{|X|c|c|c|c|}
				\hline
				\textbf{Setting} & \textbf{CT}   & \textbf{CF} & \textbf{IT}  & \textbf{Score} \\
				\hline 
                Baseline  & 1832    & 997   & 3   & 4565   \\ \hline
                Optimization Only  & 1934    & 983   & 3  & 4755     \\ \hline
                All Instructions Full  & 1925   & 926  & 3  & 4680  \\ \hline
                All Instructions Local  & 1932   & 990  & 3   & 4758 \\ \hline
                Guard Instructions Full  & 1928   & 988  & 3   & 4748 \\ \hline
                Guard Instructions Local  & 1931   & 986  & 3  & 4752 \\ \hline
                Loop Instructions  & 1989    & 987   & 3  & 4869 \\ \hline
			\end{tabularx}
\end{table}

\begin{table}[t]
	\caption{Interval precision and representation domain ($\mathbb{I}$ - integer, $\mathbb{W}$ - wrapped).
 } \label{table:precision-results} 
			\begin{tabularx}{\textwidth}{|X|c|c|c|c|c|c|c|c|}
				\hline
				\textbf{Setting} & \textbf{CT-$\mathbb{I}$} & \textbf{CT-$\mathbb{W}$}  & \textbf{CF-$\mathbb{I}$} & \textbf{CF-$\mathbb{W}$} & \textbf{IT-$\mathbb{I}$} & \textbf{IT-$\mathbb{W}$} & \textbf{Score-$\mathbb{I}$} & \textbf{Score-$\mathbb{W}$}  \\
				\hline 
                No Arithmetic  & 1991  & 1942  & 988 & 921  & 3  & 3 & 4874  & 4709 \\ \hline
                Arithmetic  & 1640  & 1640  & 709 & 519  & 0  & 0 & 3989  & 3799 \\ \hline 
                Arithmetic \& Widening  & 1992  & 1896  & 988 & 717  & 3  & 3 & 4876  & 4413 \\ \hline
			\end{tabularx}
\end{table}

\subsubsection{Precision} In Table~\ref{table:precision-results}, we demonstrate the impact of enabling the Arithmetic and Widening operators in conjunction with the optimizations and Loop Instructions setting. The results show that enabling Arithmetic causes around 30\% more timeouts. This is expected as the additional precision of Arithmetic requires more iterations to reach a fixed-point. Enabling Widening makes the issue disappear. 
Introducing more precise interval computation does not reduce the SMT solving time (<1\% impact). However, the Arithmetic setting can solve 3 unique ECA benchmarks by finding a violation in less than 90 seconds.

\subsubsection{Representation} Also in Table~\ref{table:precision-results}, we compare the use of the Integer and Wrapped domains. 
The results show that ESBMC solves fewer benchmarks with the Wrapped domain. This is because the Wrapped domain has an extra cost associated with the Join operation, thus resulting in more timeouts. This effect is most visible in the Hardware category, where the time needed to verify the benchmarks doubled.

\subsection{Limitations}

Although we successfully demonstrated the performance of our prototype on $9538$ different benchmarks, there are a few known limitations in our implementation: 

\begin{itemize}
    \item We provide no support for dereferencing operations. Any storage operation over dereferences resets the entire domain to \textit{supremum} to maintain soundness. Adding such support might improve the precision of interval analysis.


\item Programs that contain function pointers may lead to incorrect invariants. This is because the ESBMC memory model has inconsistent behavior when dealing with values being cast into pointers.\footnote{\url{https://github.com/esbmc/esbmc/issues/1539}}
\end{itemize}

\section{Conclusions}
\label{sec:conclusions}

This work presents a case study using interval analysis in a BMC framework for software verification. Specifically, we compare interval domains supporting infinite integer precision and machine-aware representations. Furthermore, we conducted a series of experiments with different levels of interval precision and analyzed their impact on BMC program analysis. 

Our results show that using intervals to optimize and instrument the program increases the SV-COMP scores from $6655$ to $7008$, with $202$ unique benchmarks solved. Additionally, using intervals in real-world software enabled our prototype to verify it. Additionally, interval analysis enables us to verify large-scale real-world firmware. Indeed, the only downside of interval analysis is its additional computational cost, which can lead to resource exhaustion in a limited number of cases. 
In all other scenarios, our results show that a lightweight interval analysis with low precision yields stronger invariants, which can be exploited by the BMC and $k$-induction algorithms.

In the future, we plan to expand the current work in two directions. First, we will add support for other types of assertions (i.e., overflows and memory safety), enabling us to infer stronger interval invariants. Second, we will apply interval analysis during symbolic execution (like in~\cite{4527113}), which will allow us to reason on a bounded trace.


%
%

\section*{Acknowledgements}

This work is partially funded by ARM, EPSRC EP/T026995/1, EPSRC EP/V000497/1, Ethereum Foundation, EU H2020 ELEGANT 957286, UKRI Soteria, Intel, and Motorola Mobility (through Agreement N° 4/2021).


\bibliographystyle{splncs04}
\bibliography{main}

\appendix

\section{GOTO Language}
\label{app:algorithms}

A program written in the language evaluates to \emph{true} if there are no inputs that can cause an assertion failure. For this work, the language has the following characteristics:
\begin{itemize}
    \item No support for pointers;
    \item The values are typed by width and sign.
    \item No recursive expressions. All expressions have an intermediate variable to hold the 
          values: $x + y + 1$ becomes $t = x + y \rightarrow t + 1$.
    \item The complete relation $P_t$ that maps all program variables into types: $\mathbb{V} \rightarrow \mathbb{Ty}$ such that $\forall v \in  \mathbb{V}$, $\ P_t(v) \in \mathbb{Ty}$. In other words, this is similar to a compiler symbol table~\cite{parsons1992introduction}, containing information of all types for every variable.
\end{itemize}

\paragraph{Basics}

We will start describing the basic features of the language. The language contains typed variables and 32-bit signed constants. The supported types are signed and unsigned integers. Additionally, the language has support for explicit program labels.

\begin{bnfgrammar}
v \in $\textbb{V}$ : \textit{Variables}
;;
n \in $\textbb{M}$ : \textit{32-bit signed integers}
;;
l \in $\textbb{L}$ : \textit{Explicit program Labels}
;;
type \in $\textbb {Ty}$ : \textit{Types} ;;
type ::= Signed n | Unsigned n
;;

term \in $\textbb{T}$ : \textit{Program terminals} ;;
term ::=
v  : variable
| n : constant 
\end{bnfgrammar}

\paragraph{Expressions}

The language has support for classic arithmetic expressions, boolean operations and bitwise operations.

\begin{bnfgrammar}
arith-expr \in $\textbb {A}$ : \textit{Arithmetic Expressions} ;;
arith-expr ::=
term $\oplus$ term : $\textbb {\forall \oplus \in \{+. -, *, / \} }$
\end{bnfgrammar}

\begin{bnfgrammar}
bool-expr \in $\textbb {B}$ : \textit{Boolean Expressions} ;;
bool-expr ::=
term \&\& term : conjuction
| !term: negation
| term $\leq$ term : less than equal
| term : plain value
\end{bnfgrammar}

\begin{bnfgrammar}
bitwise-expr \in $\textbb {Bw}$ : \textit{Bitwise Expressions} ;;
bitwise-expr ::= (type) term : casting
| $\sim$term : bitflip
| term $\oplus$ term : $\textbb {\forall \oplus \in \{<<. >>, \&, \|, \hat{} \} }$
\end{bnfgrammar}

\begin{bnfgrammar}
expr \in $\textbb {E}$ : \textit{Expressions} ;;
expr ::= arith-expr : arithmetic expression
| bool-expr  : boolean expression
| bitwise-expr : bitwise expression
\end{bnfgrammar}

\subsection{Semantics of GOTO Language}

\subsubsection{Term Semantics}

In order to define values for the terms, we need to get the values for all the variables. For that, we first define the relation $\mathbb{Ev}$ that maps $\mathbb{V} \rightarrow \mathbb{M}$. An environment is a $\rho \in \mathbb{Ev}$, such that for a $v \in  \mathbb{V}$, $\rho(v) \in \mathbb{M}$. Given an environment $\rho \in (\mathbb{V} \rightarrow \mathbb{M})$, and an $T \in \mathbb{T}$. Let $\mathcal{T} \llbracket T \rrbracket \rho$ be the computation of the term into $\mathbb{M}$, defined as: 

\begin{align*}
    \mathcal{T} \llbracket i \in \mathbb{Z} \rrbracket \rho &\triangleq i \\
    \mathcal{T} \llbracket v \in \mathbb{V} \rrbracket \rho &\triangleq \rho(v) 
\end{align*}

\subsubsection{Expression Semantics}

When computing expressions, we use the same semantics of C99 standard (including the implicit casts). The arithmetic expressions can \emph{overlap} (unsigned) or \emph{overflow} (signed). For overlaps, we use the same semantics shown by the C99 standard. For overflow, the semantics are undefined. Given an environment $\rho \in (\mathbb{V} \rightarrow \mathbb{M})$, and an $A \in \mathbb{A}$. Let $\mathcal{A} \llbracket A \rrbracket \rho$ be the computation of the arithmetic expression into $\mathbb{Z}$, let $\mathcal{B}_w \llbracket B_w \rrbracket \rho$ be the computation of the bitwise expression into $\mathbb{M}$, defined as: 

\begin{align*}
    \mathcal{A} \llbracket t_0 \oplus t_1 \rrbracket \rho &\triangleq (\mathcal{T} \llbracket t_0 \rrbracket \rho \oplus \mathcal{T} \llbracket t_1 \rrbracket \rho), \forall \oplus \in \{+, -, *, /\}    
\end{align*}

Semantics for the Bitwise operations $\mathcal{B}_w \llbracket B_w \rrbracket \rho$ and boolean expressions $\mathcal{B} \llbracket B \rrbracket \rho$ are defined in similar fashion. Finally, given an environment $\rho \in (\mathbb{V} \rightarrow \mathbb{M})$, and an $E \in \mathbb{E}$. Let $\mathcal{E} \llbracket E \rrbracket \rho$ be the computation of an expression $E$ into $\mathbb{M}$, defined as: 

\begin{align*}
    \mathcal{E} \llbracket e \in \mathbb{A} \rrbracket \rho &\triangleq \mathcal{A} \llbracket e \rrbracket \rho  \\
    \mathcal{E} \llbracket e \in \mathbb{B} \rrbracket \rho &\triangleq \mathcal{B} \llbracket e \rrbracket \rho  \\
    \mathcal{E} \llbracket e \in \mathbb{Bw} \rrbracket \rho &\triangleq \mathcal{B}_w \llbracket e \rrbracket \rho  
\end{align*}

\subsubsection{Statement Semantics}

Given an environment $\rho \in (\mathbb{V} \rightarrow \mathbb{Z})$, and an $S \in \mathbb{S}$. Let $\mathcal{S} \llbracket S \rrbracket \rho$ be the computation of an expression $S$ into $\mathbb{Ev}$, defined as: 

\begin{align*}
    \mathcal{S} \llbracket s_1;s_2 \rrbracket \rho &\triangleq \mathcal{S} \llbracket s_2 \rrbracket(\mathcal{S} \llbracket s_1 \rrbracket \rho)  \\
    \mathcal{S} \llbracket \text{Assignment v e} \rrbracket \rho &\triangleq \forall x \in \mathbb{V}. x = v \rightarrow \mathcal{A} \llbracket e \rrbracket \rho ; x \neq v \rightarrow \rho(x) \\ 
    \mathcal{S} \llbracket s \in \mathbb{S} \rrbracket \rho &\triangleq \rho  \\
\end{align*}


\section{Interval bitwise operators with integers of any bit width}
\label{app:warren}

In Section \ref{sec:goto_intervals} we presented our interval analysis framework for 32-bit signed integer variables. In reality, C code supports signed and unsigned integers of various bit length (typically 8, 16, 32 and 64). As a result, we need to define interval-based operators for these integers as well.

While arithmetic operations are a trivial extension of Table \ref{table:integer-operators}, bitwise operations require a little more care. For 32-bit integers, the work of \cite{warren2013hacker} provide interval algorithms that cover the operators OR, AND, XOR and NOT. Integers of length different than 32 bits are not covered, as well as shift and typecast operators (truncation, extension).

\begin{table}[t]
	\caption{ Operators for Integer Domain.} \label{table:integer-operators}
			\begin{tabularx}{\textwidth}{|X|X|}
				\hline
				\textbf{Operand} & \textbf{Result}  \\
                \hline
                $[l_0, u_0] +_I + [l_1, u_1]$ & $[l_0 + l_1, u_0 + u_1]$  \\
				\hline
                $[l_0, u_0] -_I + [l_1, u_1]$ & $[l_0 - u_1, u_0 - l_1]$  \\
				\hline
                $[l_0, u_0] *_I + [l_1, u_1]$ & $[min(l_0 * u_0, l_0 * u_1, l_1 * u_0, l_1 * u_1), max(l_0 * u_0, l_0 * u_1, l_1 * u_0, l_1 * u_1)]$  \\
				\hline
                $[1, 1] /_I [l, u]$ & $ u < 0 \vee 0 < l \rightarrow [1/u, 1/l]; \textit{otherwise} \rightarrow (-\infty, +\infty)$  \\
				\hline
                 $[l_0, u_0] /_I [l_1, u_1]$ & $[l_0, u_0] * ([1,1] / [l_1, u_1])$  \\
				\hline
			\end{tabularx}
\end{table}

Here, we present our bit-agnostic implementation of bitwise operators. These can be used with integers of any length (not necessarily powers of two), and offer a considerable performance improvement over the algorithms in \cite{warren2013hacker}.

The code for computing the lower (respectively upper) bounds of a bitwise or operation $[$minOR$,$maxOR$]=[a,b]\:|\:[b,c]$ is given in Listings \ref{lst:min_or} and \ref{lst:max_or}. For ease of comparison, we present the 32-bit algorithm in \cite{warren2013hacker} side-by-side. Note how the algorithm in \cite{warren2013hacker} requires scanning all 32 bits via the one-hot variable \texttt{m}, from most significant to least significant. In contrast, our algorithms have two advantages. First, they perform the scan from least to most significant bit via the one-hot variable \texttt{lsb}. As a result, we can support \texttt{unsigned} integers of any bit length. Second, we only consider the bits that are important for the final result, thus executing the body of the \texttt{while(m != 0)} loop for fewer iterations than the bit length of \texttt{unsigned}.

\begin{figure}[t]
\begin{tcolorbox}[sidebyside]

\centering
  \begin{minted}{c}
uint32_t minOR(uint32_t a,
               uint32_t b,
               uint32_t c,
               uint32_t d) {
  uint32_t m, tmp;
  m = 0x80000000;
  while (m != 0) {
    if (~a & c & m) {
      tmp = (a | m) & -m;
      if (tmp <= b) {
        a = tmp;
        break;
      }
    }
    else if (a & ~c & m) {
      tmp = (c | m) & -m;
      if (tmp <= d) {
        c = tmp;
        break;
      }
    }
    m = m >> 1;
  }
  return a | c;
}
  \end{minted}
\tcblower

\begin{minted}{c}
unsigned bestOR(unsigned x,
                unsigned y,
                unsigned m) {
  unsigned best, lsb, tmp;
  best = x;
  while(m != 0) {
    lsb = m & (-m);
    tmp = (x | lsb) & (-lsb);
    if(tmp > y) break;
    best = tmp;
    m &= (m - 1);
  }
  return best;
}

unsigned minOR(unsigned a,
               unsigned b,
               unsigned c,
               unsigned d) {
  unsigned best_a, best_c;
  best_a = bestOR(a, b, ~a & c);
  best_c = bestOR(c, d, a & ~c);
  unsigned m = best_a | c;
  unsigned n = a | best_c;
  return (m < n)? m: n;
}
  \end{minted}

\end{tcolorbox}
 \caption{Left: original algorithm for 32-bit unsigned or (lower bound) in \protect\cite{warren2013hacker} page 75. Right: optimised version for integers of any bit length.} \label{lst:min_or}
\end{figure}

The other bitwise operators (signed and unsigned AND, OR, XOR and typecasts) can be defined in terms of \texttt{minOR} and \texttt{maxOR}. Furthermore, shifts and negation require few lines of code. For more details on these algorithms, refer to the file \texttt{bitwise\_bounds.h} in our GitHub repository\footnote{\url{https://github.com/esbmc/esbmc/blob/master/src/goto-programs/abstract-interpretation/bitwise\_bounds.h}} (or in the Zenodo artifact) and the commented examples therein.

\begin{figure}[t]
\begin{tcolorbox}[sidebyside]

\centering
  \begin{minted}{c}
uint32_t maxOR(uint32_t a,
               uint32_t b,
               uint32_t c,
               uint32_t d) {
  uint32_t m, tmp;
  m = 0x80000000;
  while (m != 0) {
    if (b & d & m) {
      tmp = (b - m) | (m - 1);
      if (tmp >= a) {
        b = tmp; 
        break;
      }
      tmp = (d - m) | (m - 1);
      if (tmp >= c) {
        d = tmp;
        break;
      }
    }
    m = m >> 1;
  }
  return b | d;
}
  \end{minted}
\tcblower

\begin{minted}{c}

unsigned maxOR(unsigned a,
               unsigned b,
               unsigned c,
               unsigned d) {
  unsigned e, m, lsb;
  unsigned tmp_b, tmp_d;
  e = 0;
  m = b & d;
  while(m != 0) {
    lsb = m & (-m);
    tmp_b = (b - lsb) | (lsb - 1);
    tmp_d = (d - lsb) | (lsb - 1);
    if(tmp_b < a && tmp_d < c)
      break;
    e |= (lsb - 1);
    m &= (m - 1);
  }
  return b | d | e;
}
  \end{minted}

\end{tcolorbox}
 \caption{Left: original algorithm for 32-bit unsigned or (upper bound) in \cite{warren2013hacker} page 77. Right: optimised version for integers of any bit length.}
 \label{lst:max_or}
\end{figure}

\section{ESBMC Details}
\label{app:esbmc-options}

In this Appendix, we will present selected options of ESBMC that are used throughout this work. In Table~\ref{table:esbmc-flags}, we show the flags and their explanations. Although the interval domain and precision can be selected through flags, the instrumentation and optimization can not (we provide ESBMC builds for each instrumentation/optimization configuration in Zenodo). 

\subsubsection{Industrial Example (Section~\ref{sec:firmware}).}
    --no-pointer-check --no-div-by-zero-check --no-bounds-check --no-vla-size-check --no-align-check --no-pointer-relation-check --no-unlimited-scanf-check --unwind 1 --partial-loops --interval-analysis

\subsubsection{SV-COMP Baseline (Section~\ref{sec:SV-COMP}).}

--no-div-by-zero-check --force-malloc-success --state-hashing --add-symex-value-sets --no-align-check --k-step 2 --floatbv --unlimited-k-steps --no-vla-size-check --32/--64 --witness-output witness.graphml --enable-unreachability-intrinsic --no-pointer-check --no-bounds-check --error-label ERROR --goto-unwind --unlimited-goto-unwind --k-induction --max-inductive-step 3

\subsubsection{SV-COMP Intervals (Section~\ref{sec:SV-COMP}).} --no-div-by-zero-check --force-malloc-success --state-hashing  --add-symex-value-sets --no-align-check --k-step 2 --floatbv --unlimited-k-steps --no-vla-size-check --32/--64  --witness-output witness.graphml --enable-unreachability-intrinsic --no-pointer-check  --no-bounds-check --error-label ERROR --goto-unwind --unlimited-goto-unwind --k-induction --max-inductive-step 3 --interval-analysis

\subsubsection{Application unique (Section~\ref{sec:invariant-mode}).} --no-div-by-zero-check --force-malloc-success --state-hashing  --add-symex-value-sets --no-align-check --k-step 2 --floatbv --unlimited-k-steps --no-vla-size-check --32/--64  --witness-output witness.graphml --enable-unreachability-intrinsic --no-pointer-check  --no-bounds-check --error-label ERROR --goto-unwind --unlimited-goto-unwind --k-induction --max-inductive-step 3 --interval-analysis

\subsubsection{Precision unique (Section~\ref{sec:invariant-mode}).} --no-div-by-zero-check --force-malloc-success --state-hashing  --add-symex-value-sets --no-align-check --k-step 2 --floatbv --unlimited-k-steps --no-vla-size-check --32/--64  --witness-output witness.graphml --enable-unreachability-intrinsic --no-pointer-check  --no-bounds-check --error-label ERROR --goto-unwind --unlimited-goto-unwind --k-induction --max-inductive-step 3 --interval-analysis --interval-analysis-bitwise? --interval-analysis-arithmetic? --interval-analysis-extrapolate?

\subsubsection{Representation unique (Section~\ref{sec:invariant-mode}).} --no-div-by-zero-check --force-malloc-success --state-hashing  --add-symex-value-sets --no-align-check --k-step 2 --floatbv --unlimited-k-steps --no-vla-size-check --32/--64  --witness-output witness.graphml --enable-unreachability-intrinsic --no-pointer-check  --no-bounds-check --error-label ERROR --goto-unwind --unlimited-goto-unwind --k-induction --max-inductive-step 3 --interval-analysis --interval-analysis-bitwise? --interval-analysis-arithmetic? --interval-analysis-extrapolate? --interval-analysis-wrapped


\begin{table}[h]
	\caption{ESBMC flags and descriptions.} \label{table:esbmc-flags}
			\begin{tabularx}{\textwidth}{|c|X|}
				\hline
				\textbf{Flag} & \textbf{Description}  \\
                \hline
                \texttt{--floatbv} & Encode floats using BitVector theory  \\
				\hline
                \texttt{--k-induction} & Enables the incremental k-induction strategy  \\
				\hline
                \texttt{--unlimited-k-steps} & Remove any upper bound for the BMC incremental approach  \\
				\hline
                \texttt{--k-step 2} & Increments each bound limit by 2 for iteration  \\
				\hline
                \texttt{--max-inductive-step 3} & Do not apply the inductive step over bound 3  \\
				\hline                
                \texttt{--32/--64} & Sets the architecture  \\
				\hline
                \texttt{--no-div-by-zero-check} & Disable division by zero assertions \\
				\hline
                \texttt{--no-align-check} & Disable memory alignment check \\
				\hline
                \texttt{--no-vla-size-check} & Disable out-of-bounds check over VLAs  \\
				\hline
                \texttt{--no-pointer-check} & Disable pointer safety checks  \\
				\hline
                \texttt{--no-pointer-relation-check} & Disable pointer relation checks   \\
				\hline
                \texttt{--no-unlimited-scanf-check} &  Allow partial check over scanf inputs \\
				\hline
                \texttt{--no-bounds-check} & Disable out-of-bounds checks  \\
				\hline
                \texttt{--state-hashing} & Removes duplicate states in interleaving (for concurrency)  \\
				\hline
                \texttt{--add-symex-value-sets} & Adds assumptions of pointer destinations (during symbolic execution)    \\
				\hline
                \texttt{--goto-unwind} & Statically unroll loops of known lengths    \\
				\hline
                \texttt{--enable-unreachability-intrinsic} & Allows early termination for some float point operational models  \\
				\hline
                \texttt{--partial-loops} &  allows partial execution of loops   \\
				\hline
                \texttt{--force-malloc-success} &  assumes that dynamic allocation can never fail   \\
				\hline
                \texttt{--unlimited-goto-unwind} & Removes the upper bound limit from \texttt{goto-unwind}\\
                \hline
                \texttt{--witness-output} &  construct a witness report using SV-COMP format   \\
				\hline 
                \texttt{--error-label} &  assumes that reaching the defined program label is an error   \\
				\hline 
                \texttt{--interval-analysis} & Enables the use of Interval Analysis  \\ \hline
                \texttt{--interval-analysis-wrapped} & Replaces Integer Domain with Wrapped Domain  \\
				\hline
                \texttt{--interval-analysis-arithmetic} & Enables arithmetic computations during interval analysis  \\ \hline
                \texttt{--interval-analysis-bitwise} & Enables bitwise computations during interval analysis  \\ \hline
                \texttt{--interval-analysis-extrapolate} & Enables widening during interval analysis  \\ \hline
			\end{tabularx}
\end{table}


\end{document}